\documentclass[sigconf,authorversion]{acmart}

\usepackage{bm}
\usepackage[english]{babel} 
\usepackage{underscore}
\usepackage{multirow}
\usepackage{float}
\usepackage{amsmath}
\usepackage[export]{adjustbox}
\usepackage{booktabs}

\usepackage{enumitem,kantlipsum}
\AtBeginDocument{%
  \providecommand\BibTeX{{%
    \normalfont B\kern-0.5em{\scshape i\kern-0.25em b}\kern-0.8em\TeX}}}

\acmConference[SIGIR '23]{the 46th International ACM SIGIR Conference on Research and Development in Information Retrieval (SIGIR '23)}{July 23-27, 2023}{Taipei, Taiwan}
\begin{document}

\title{Exploring  360-Degree View of Customers  for Lookalike Modeling}

\author{
Md Mostafizur Rahman$^{1}$,
Daisuke Kikuta$^{1*}$,
Satyen Abrol$^{2}$,
Yu Hirate$^{1}$, 
Toyotaro Suzumura$^{1, 3}$,
Pablo Loyola$^{1}$,
Takuma Ebisu$^{1}$,
Manoj Kondapaka$^{2}$
}

\thanks{$^*$The author was with Rakuten Institute of Technology when the work was done}
\affiliation{%
 \vspace{2.5mm}$^{1}$Rakuten Institute of Technology, Rakuten Group, Inc., \country{Japan}
}
\affiliation{%
 $^{2}$Rakuten Institute of Technology, Rakuten Group, Inc., Bengaluru, India
 \country{Japan}
}
\affiliation{%
 $^{3}$The University of Tokyo, Tokyo, Japan
  \country{Japan}
}

\renewcommand{\shortauthors}{Rahman et al.}

\begin{abstract}
Lookalike models are based on the assumption that user similarity plays an important role towards product selling and enhancing the existing advertising campaigns from a very large user base. Challenges associated to these models reside on the heterogeneity of the user base and its sparsity. 
In this work, we propose a novel framework that unifies the customers’ different behaviors or features such as demographics, buying behaviors on different platforms, customer loyalty behaviors and build a lookalike model to improve customer targeting for Rakuten Group, Inc. Extensive experiments on  real e-commerce and travel datasets demonstrate the effectiveness of our proposed lookalike model for user targeting task.
\end{abstract}


\maketitle

\section{Introduction}

E-commerce and online businesses require efficient lookalike models to recommend the right products/services to the right customer. Tech companies e.g., Facebook, Google \cite{kanagal2013focused}, LinkedIn \cite{liu2016audience} provide several marketing campaign platforms. Advertisers and marketers often seek to target users similar to the users who already showed interest on their products to maximize the likelihood of positive responses to their user targeting. Audience expansion, also called look-alike modeling, is the task of finding users who are similar to a given set of seed users and likely to achieve the business goal of the target product/campaign. The seed users can be, for example, purchasers of a product, web subscribers, or loyal customers.  

Any lookalike model suffers from several challenges e.g., features selection or depending on only demographic or buying behaviors; limited seed users which may lead to overfitting; each target prospecting task is different and can cover almost all the major topics. To address those challenges, we propose Embedding based Customer Lookalike Modeling (E-CLM), which utilizes a customer 360-degree view. E-CLM relies on the customer personality, buying behavior, cross service data and loyalty information.

Rakuten has  one of the largest e-commerce platforms in the world known as Rakuten Ichiba. In addition to e-commerce service, Rakuten provides more than 90 different online services and Rakuten Travel is one of them. We observed that a significant number of customers are active in our Ichiba (e-commerce) service while inactive in Rakuten travel service and vice versa.
In this work, we combine Rakuten Ichiba and Rakuten Travel service to predict the lookalike customers. We believe that a lookalike model should not only focus on customers’ preference but also other factors like affordability, and lifestyle. As an example, one customer can be very active on our travel service and regularly browsing/booking luxurious hotels but never active on our Ichiba service. There is a chance, that the customer can afford expensive cosmetics items but has never been targeted while lookalike model was using only one service related data. To address such issues, we include users' cross service behaviors, loyalty, personal and family views in our proposed model. 

Rakuten AIris Target Prospecting (AIris TP) is a framework which finds prospective customers for Rakuten's businesses and its clients. This framework has been deployed in Rakuten for customer targeting since 2018. 
Therefore, we present the best performing models on the AIris TP framework  in this work. In E-CLM, we propose five different views of customers: demography, loyalty, ichiba (e-commerce), travel and family view, and we learn a user representation from each of them. 
We apply E-CLM to the AIris TP Framework and evaluate the effectiveness. We also discuss another extension  of E-CLM (E-CLM++)   to improve its current  performance. 
The main contributions of our work can be stated as (a) To the best of our knowledge, we present the first lookalike model which defines the different views of customers explicitly and combine them to better learn user representation which leads to effective lookalike modeling, and (b) Comprehensive experiments to demonstrate our approach’s effectiveness as it achieves a significant accuracy comparable to the baselines set for the AIris TP.

\section{Related Work}
In this section, we introduce the related work from various aspects. Some work uses similarity-based methods \cite{ ma2016score} or clustering \cite{ wang2010discovering, cavallo2018clustrophile, jiang2019comprehensive, wenskovitch2017towards, zhou2008energy}, meta-task learning \cite{ li2018learning, zhu2021learning, chan2021interactive, xi2021modeling}
, matrix factorization \cite{ kanagal2013focused} ,  classification-based methods \cite{ bagherjeiran2010large} and rule-based methods \cite{ shen2015effective} . Knowledge graphs embedding models \cite{ RESCAL, TransE, ComplEX,DistMult}and GNNs \cite{GCN, GraphSAGE, GAT,KGCN, RGCN, HGT,zhang2019multi,liu2020exploring} based models also can be used in recommender systems to address lookalike modeling. There has also been some work that extracts rules/feature vectors for scoring. Similarity-based methods expand a given seed-list via calculating the similarity of all pairs between seed users and candidate users. Rule-based methods match similar users with specific demographic features or interests or personas that are targeted by marketers. 
LinkedIn researchers  proposed a campaign-agnostic and campaign-aware lookalike expansion approach to expand user profiles to mitigate the coverage problem and showed that while advertisers cannot  cover all relevant demographic attributes related to the desired product or service on such cases their model can achieve high accuracy \cite{liu2016audience}. However, none of related works use cross service data and explicitly encounter all customer perspectives.
\section{Embedding based Customer Lookalike Modeling}

\subsection{Problem Statement and Preliminary}
In a lookalike setting, a list of $m$ seed users $S_u={(u_1,u_2,\dots,u_m)}$ is given to the model and the task is to find $n$ similar users to the seed list $S_u$ where $n \gg m$. 

We have constructed two different knowledge graphs: one is based on e-commerce(Ichiba) service and another is based on travel service. For notations, we denote a triplet by $(u_i, r, u_j)$. Demography  and loyalty attributes are denoted as $a_d$ and $a_l$ respectively. We use bold lowercase letters to represent embeddings and bold uppercase letters for matrices. 

\subsection{Demography View (D. view)}
We observed that user demography is very important features for lookalike modeling. In our datasets, all the demography information are stored in literal/numeric format (age, reg\_date, area\_code). We represent the user demography information in triplet format $(u_d,r_d,a_d)$ where $r_d$ is the demography relation for user $u_d$.    
We employ the basic TransE \cite{TransE} for attributes and literals embedding and we interpret relation $r_d$ as a translation from the head entity $u$ to the literal $l$. 

Since the literals values can be in different formats. For example, user, $u_d$ is 55.5 years old, the area code she lives in is 14, the registration date to the e-commerce service is  '2005-12-04'. Therefore, to encode the literal value, we use a compositional function $\phi(a_d)$ where $a_d$ is a sequence of the characters of the demography attribute value $l = {c_1, c_2, c_3, \dots, c_k}$, where $c_1$, $c_2$, \dots $c_k$ are the character embeddings of the demography attribute value and define the relationship of each element in literal triple as $u_d + r_d \approx \phi(a_d)$. 
We employ an N-gram-based compositional function in our proposed E-CLM model. We use the summation of n-gram combination of the attribute value. The equation is: $\phi(a_d)=\sum_{n=1}^N   \left({\sum_{i=1}^{k}\sum_{j=i}^{n}c_j}/{(k-i-1)} \right)$.

where $N$ indicates the maximum value of n used in the ngram combinations  and $k$ is the
length of the attribute value.

To learn the demography view embeddings, we minimize the following objective function $L_{D}$:
\begin{multline}
L_{D}=\sum_{{(u_d,r_d,a_d)} \in {T_{a_d}}}\sum_{{(u_d',r_d,a_d')} \in {T'_{a_d}}} max(0, \gamma + \alpha(f_d(u_d,r_d,a_d) \\-f_d(u_d',r_d',a_d')))
\end{multline}

\noindent Here, $T_{a_d}$ is the set of valid demography attribute triples from the training dataset, while $T^{'}_{a_d}$ is the set of negative demography attribute triples. 
Here, $f_l(u_d,a_d)$ is the plausibility score based on the embedding of the head entity $u_d$, the embedding of the relationship $r_d$, and the vector representation of the demography attribute value that computed using the compositional function $\phi(a_d)$.\\

\subsection{Loyalty View (L. view)}

On our website, items are displayed under different shops or merchants. Users show strong preference to specific shops/merchants, genres, hotels or brands. It is obvious that users browse multiple e-commerce websites to compare product prices/services and each e-commerce company provides a different set of coupons or discounts for the product. So, we are interested to know about our customer/user loyalty information when they buy from our websites.  To identify the loyalty embeddings we focus on users shop, genre, hotel and targeted brand loyalty attribute. If a user buys an item from a shop more than five times, we include that shop name under user shop loyalty attributes. That means one user can be loyal to multiple shops, genres and so on.  To capture the loyalty embeddings, we utilize a simple convolutional neural network (CNN) to extract features from the loyalty attributes. We represent the loyalty attribute and its value in a matrix form $ \langle r_l; v_l \rangle \in \mathbb{R}^{(2 \times d)}$, where $r_l$ is loyalty attribute and $a_l$ is the value and  feed it to a CNN to obtain the vector representation \cite{zhang2019multi}. The score function would be $f_l(u_l,r_l,v_l) = -\|{u}_l - CNN (\langle r_l; v_l \rangle)\|$ and minimize the following objective function,  $L_{L}=\sum_{{(u_l,r_l,a_l)} \in {L}} \log(1+\exp(- f_l(u_l,r_l,v_l)))$. Here, $L$ denotes the set of real loyalty attribute triples and $u_l$ is the user's loyalty view embedding.

\subsection{Structure Embedding (SE)}
Structure embeddings encode the user-user, user-item and item-item interactions by using knowledge graph embedding models. As we mentioned earlier in this study, we have built two knowledge graphs from the e-commerce (Ichiba) and travel service data. In our e-commerce KG, we have 4 entity types (user, shop, item, genre) and 12 types of edges (bought, clicked,sold\_by,item\_under\_leaf\_genre, spouse, parent, child and 5 types of genre hierarchy based relations). Travel graph consists of 4 types of entities/nodes: user, hotel\_id, month (visiting month) and reservation\_type, reservation\_partner (clients can book using different travel agency or apps) and booked, visiting\_month, reserved\_under\_partner\_id, user\_reservation\_type (there are 6 reservation types). Here we compute three different views: (1) Ichiba view, (2) travel view and (3) users' family members view. For this three views, SE has been used. In this work, we employ a translation-based KGE model and we can formalize the loss function as follows: 
\begin{align}
\label{eq:loss_kge}
  L_{\textrm{KGE}} &= \sum_{(h, r, t)\in\mathcal{E}}\sum_{(h', r, t')\in\mathcal{E}^{-1}}\left[\gamma + f(h,r,t) - f(h',r',t')\right]_+,
\end{align}
where $(h,r,t)$ is positive triples, which actually exist in the KG,  $(h',r,t')$ is broken triples, $\gamma$ is the margin, $f(h,r,t)$ is the score function, and $[\cdot]_+=max(0, \cdot)$. 
Here the score function is defined as $ f(h,r,t) = \|\bm{u}_h + \bm{u}_r - \bm{u}_t \|_{1}$, 
where $\bm{u}_h$, $\bm{u}_r$ and $\bm{u}_t$, are the embeddings of head entities, relations and tail entities, respectively. We employ the simple KGE model in our structure embedding for E-CLM because we have not seen any significant change while using ComplEX or Graph Sage. 


\begin{table}[b!]
    \scriptsize	
    \caption{Datasets statistics (A$\sim$E represent the top 5 brands from Rakuten Ichiba).}
    \label{table:airis_datasets}
    \begin{tabular}{lcccccccc}
      \hline
      & A & B & C  &  D & E & Resort R & Hotel H &Tencent\\
      \hline
    \# Seeds & 76254 & 96952 & 70670  & 1669 & 37413&467&6147&421961\\
    \# Training  &305017  &387805  &282677  &6673& 149649 & 1506 & 17391 &1812791\\
    \# Validation  &42157 &53093 &38261  & 853 & 20033&203& 2353&226598\\
    \# Testing     &44985 &50113 &36233  &4401 & 37513&433&5704&226600 \\
      \hline
 
    \end{tabular}
\end{table}
\begin{table*}[h]
\scriptsize	
 \caption{Model Performance (Precision and PR-AUC) on Rakuten Ichiba, Rakuten Travel and public dataset}
    \label{table:results}
\centering
\small
  \begin{tabular}{|l|l|l|l|l|l|l|l|l|l|l|l|l|l|l|l|l|}

    \hline
    \multirow{2}{*}{Models} &
      \multicolumn{2}{c|}{A} &
      \multicolumn{2}{c|}{B} &
      \multicolumn{2}{c|}{C} & 
      \multicolumn{2}{c|}{D} &
       \multicolumn{2}{c|}{E} &
        \multicolumn{2}{c|}{Resort R}&
       \multicolumn{2}{c|}{Hotel H}&
       \multicolumn{2}{c|}{Tencent}\\
    & Prec. & Pr-auc & Prec. & Pr-auc  & Prec. & Pr-auc   & Prec. & Pr-auc  & Prec. & Pr-auc & Prec. & Pr-auc & Prec. & Pr-auc & Prec. & Pr-auc \\
    \hline
      Baseline TP   &0.536& 0.705 & 0.549 & 0.716 &0.571 &0.747 & 0.638&0.751&0.565&0.738&0.414&0.437&0.306&0.379&0.289&0.581 \\
      LR        &0.511&0.656 & 0.547 & 0.699 & 0.545 &0.710& 0.617 & 0.723&0.551&0.723&0.397&0.413&0.286&0.413&0.272&0.570\\
      TP+TransE             &0.543&0.724&0.552&0.724&0.585&0.767&0.657&0.776&0.568&0.749& 0.456&0.447&0.341&0.418&0.290&0.612 \\
      TP+ComplEX    &0.548&0.726& 0.552&0.722& 0.590&0.765 &0.662&0.780&0.571&0.755 &0.460&0.450&0.345&0.420&0.298&0.621 \\
      TP+Graph Sage &0.552&0.724&0.558&0.728&0.598&0.760& 0.665&0.778&0.575&0.758 &0.463&0.452&0.350&0.419&0.302&0.642\\
      MetaHeac&0.549&0.724&0.554&0.724&0.593&0.766&0.661&0.780&0.570&0.752&0.421&0.441&0.314&0.385&0.314&\textbf{0.735}\\
      E-CLM &0.552&0.729&0.561&0.727&\textbf{0.601}&0.771&0.670&0.784&0.579&0.761& 0.461&0.482&0.348&0.421&\textbf{0.316}&\textbf{0.735}\\
      E-CLM++ &\textbf{0.557}&\textbf{0.732}&\textbf{0.564}&\textbf{0.733}&\textbf{0.605}&\textbf{0.774}&\textbf{0.679}&\textbf{0.791}&\textbf{0.586}&\textbf{0.766}&\textbf{0.466}&\textbf{0.483}&\textbf{0.382}&\textbf{0.423}& 0.314 & 0.734\\
       \hline
  \end{tabular}
\end{table*}
\subsubsection{Ichiba View (I. view)}
Ichiba view (e-commerce) aggregate users' user -item and item -item interactions from our e-commerce KG by applying KGE (using the Eq. (\ref{eq:loss_kge}) ) and obtain embedding for Ichiba view $u_e$.
\subsubsection{Travel View (T. view)}
Travel graph captures the users' travel enthusiasms  and interactions with travel products. In this graph, users' hotel/resort preference, reservation type etc. (please refer to Sect. 3.4 )  are interconnected. Here, we compute users' travel view, $u_{tv}$ using KGE. 
\subsubsection{Users' Family Members View (FM. view)}
Family and social life have an immense impact on customers' buying behaviors (Rakuten social graph contains different relationships among it's customers). They also reflect users' preferences to different products and services. Here, we only consider users' family members buying interactions only and exclude actual seed and non-seed users' interactions.  In Ichiba view we only learn user-related interactions but here we extract all the users' family member interactions. Users' family member view denoted as $u_f$.



\subsection{Finding Lookalike Customers}

We calculate five different views of each customer and combine the views of each user by defining a weight to each view. To emphasize on important views, we assign weights to view-specific entity embeddings \cite{liu2020exploring}. Let $u_{final}$ denotes the combined embedding for $u$. Without loss of generality, let
V be the number of views, and we have: $u_{final} = \sum_{i=1}^{V}w_i{u^{(i)}}$,
\noindent where $w_i$ is the weight of $u^{(i)}$, and can be calculated by:$ w_{i} = {cos(u^{(i)}, u")}/{\sum_{j=1}^{V}cos({u^{(j)},u"})}$,
where $u"$ is the average of user embeddings of $u$, i.e.,$u"=\frac{1}{V}\sum_{i=1}^{V}w_i{u^{(i)}}$.
If the embedding from one view is far away from its average embedding, it would have a lower weight. 
We use these embeddings in two ways: (1) use a similarity threshold $T$ to filter the closest customers of each seed customer and generate the new list as target prospecting (E-CLM); (2) use the user embeddings as pre-trained features for the current baseline model, which we will call E-CLM++.

\section{Experiments}

\begin{table}[h]
\scriptsize	
 \caption{Model Performance (Accuracy) on all the datasets.}
\begin{tabular}{lllllllllll} \toprule
    &Models &  A & B &  C  & D & E   & Resort R & Hotel H&Tencent    \\ \midrule
     &  Baseline TP &0.764&0.776&0.797&0.804&0.761&0.696&0.538& 0.670\\
     & LR &0.755&0.740&0.758&0.722&0.717&0.692&0.527& 0.654\\
     & TP+TransE &0.776& 	0.782& 	0.832& 	 	0.837& 	0.790 &0.746&0.698& 0.725 \\
     & TP+ComplEX &0.780&0.782&0.825&0.833&0.784&0.749&0.705& 0.727  \\
     & TP+Graph Sage &0.778&0.779&0.748&0.842&0.790&0.752&0.703& 0.729 \\
     &MetaHeac&0.776&0.785&0.831&0.841&0.796&0.749&0.543&\textbf{0.741}\\
     & E-CLM &0.775&0.788&0.834&0.841&0.788&0.777&\textbf{0.721}& 0.740\\
     & E-CLM++ &\textbf{0.783} &\textbf{0.806} &\textbf{0.835}  &\textbf{0.850} &\textbf{0.807}  &\textbf{ 0.780} &0.716&0.738\\ \midrule
  
\end{tabular}
\end{table}


\begin{table*}
\scriptsize	
 \caption{Ablation (Precision and PR-AUC) on  Rakuten Ichiba and  Travel dataset.}
    \label{table:results1}
\centering
\small
  \begin{tabular}{|l|l|l|l|l|l|l|l|l|l|l|l|l|l|l|l|l|l|l|}

    \hline
    \multirow{2}{*}{Models} &
      \multicolumn{2}{c|}{A} &
      \multicolumn{2}{c|}{B} &
      \multicolumn{2}{c|}{C} & 
      \multicolumn{2}{c|}{D} &
       \multicolumn{2}{c|}{E} &
        \multicolumn{2}{c|}{Resort R}&
       \multicolumn{2}{c|}{Hotel H}\\
    & Prec. & Pr-auc & Prec. & Pr-auc  & Prec. & Pr-auc   & Prec. & Pr-auc  & Prec. & Pr-auc & Prec. & Pr-auc & Prec. & Pr-auc \\
    \hline
      D-view   &0.347& 0.557 & 0.351 & 0.521 &0.399 &0.510 & 0.445&0.562&0.362&0.587&0.273&0.321&0.201&0.274 \\
      L-view        &0.221&0.401 & 0.270 & 0.399 & 0.211 &0.398 
      & 0.312 & 0.411&0.286&0.403&0.171&0.222&0.153&0.278\\
      I-View             &0.345&0.550&0.352&0.519&0.390&0.501&0.450&0.570&0.359&0.582& 0.230&0.301&0.199&0.203 \\
      T-view    &0.232&0.433 & 0.310 & 0.435 & 0.275 &0.448&  0.383 & 0.460&0.298&0.433&0.361&0.392&0.250&0.312 \\
      FM-view  &0.212&0.398 & 0.261 & 0.373 & 0.208 &0.388& 0.314 & 0.419&0.286&0.403&0.225&0.274&0.222&0.316\\
      D+L&0.412& 0.587 & 0.415 & 0.590 &0.459 &0.599 & 0.513&0.622&0.423&0.648&0.344&0.424&0.285&0.342\\
      I+T+FM &0.432& 0.611 & 0.416 & 0.594 &0.455 &0.586 & 0.518&0.631&0.432&0.663&0.366&0.421&0.300&0.365\\
      D+L+I &0.492& 0.689 & 0.497 & 0.671&0.541 &0.701 & 0.611&0.721&0.500&0.699&0.356&0.425&0.297&0.362\\
      D+L+T &0.400& 0.599 & 0.427 & 0.595 &0.513 &0.600 & 0.523&0.641&0.413&0.651&0.360&0.433&0.302&0.391\\
      D+L+I+T&0.540 &0.718 &0.552 &0.721 &0.594&0.760 &0.661&0.771&0.568&0.745& 0.447&0.445&0.322&0.401\\
    
      E-CLM  (all views) &0.552&0.729&0.561&0.727&0.601&0.771&0.670&0.784&0.579&0.761& 0.461&0.482&0.348&0.421\\
       \hline
  \end{tabular}
\end{table*}


\subsection{Experimental Settings}

\subsubsection{Datasets}
\label{subsec:datasets}
\begin{itemize}[leftmargin=*]

\item\textbf{Rakuten Ichiba and Travel datasets:} Ichiba dataset contains top 5 brands (by revenue generation) sold in Rakuten. For anonymity purpose code names have been used. Brand A and Brand E are very popular brands for health-related products e.g., vitamins, mineral, stress relievers etc.  Brand B and C sell beauty products and toiletries. Brand D is a famous beverage brand. In Rakuten Travel dataset , Resort R is a very famous luxury resort chain in Japan, which has resorts in four different locations. Hotel H is a standard hotel in Tokyo.  
\item \textbf{Tencent dataset} It is a public dataset\footnote{https://pan.baidu.com/s/1tzBTQqA0Q9qexFr32hFrzg (password: ujmf)} which was proposed for Tencent Ads competitions in 2018. This dataset contains hundred of different seed sets (total 421,961 seed users). In this dataset, each user has 14 features including  user demography and insterests. Each advertisment task in this dataset has 6 categorical features: ad category, advertiser ID, campaign ID, product ID, product type, creative size. We can understand that this dataset has no cross domain information and features are quite different from Rakuten Ichiba and Travel datasets. For the Tencent dataset, we have prepared two views for our proposed E-CLM models. Demography and Interest view: This view has been calculated using demography and interest features of this dataset in a similar way, as described in Sect. 3.2; Tencent structured view: Similar to Ichiba view, we aggregate user to product and capture the connections with product to other categorical features in a KG format, and obtain Tencent KG view.\par 
\end{itemize}

The details of the datasets are shown in Table \ref{table:airis_datasets}. Please note, for Rakuten top clients we run the campaigns in monthly/bi-monthly basis. For the all datasets, the ratio of seed and non-seed users is 1:3. Each dataset contains a seed-list  and a non-seed list for training and validation, and a set of expanded customers for testing. 
\subsubsection{Baselines} 
\begin{itemize}[leftmargin=*]
\item \textbf{Baseline Target Prospecting (TP)}:  The current baseline model for AIris TP is based on XGBoost model. This model uses four different features for Rakuten datsets:
 \textit{(i) Demographic Features}: Demographic features such as age, gender, region.
 \textit{(i) Points Summary}: Rakuten users can gain points while buying different products/services. This feature exhibits point status such as current available points. 
 \textit{(iii) Point Features}: Transaction of points such as acquired/used points from online/offline shops/ merchants. 
 \textit{(iv) Genre level Purchase History}: Like other E-commerce companies, Rakuten Group maintains "genre" hierarchy. In this feature we capture shopping trends in popular genres. For the Tencent dataset all features are directly fed to a XGBoost model.
\item\textbf{Logistic Regression-based Lookalike Model (LR):} In this model \cite{qu2014systems}, for a given user $u_i$ who has a feature vector $v_i$, a logistic regression model gets trained and model the probability of being lookalike users to seed users. 
 
 \item\textbf{TP+TransE, TP+ComplEX, TP+Graph Sage}: In TP+TransE, TransE \cite{TransE} generated vectors (from the structure embedding using e-commerce KG) have been exploited as pre-trained features to Baseline TP for user expansion as baseline. In the same way as "TP+TransE" we  prepared two other baselines for our E-CLM model using ComplEX \cite{ComplEX} and graph sage \cite{GraphSAGE} .
  
    \item\textbf{MetaHeac}: It is a state-of-the-art lookalike model and has been deployed in WeChat \cite{zhu2021learning}.

\end{itemize}

\subsubsection{Model Settings}
The models run on a single GPU 'NVIDIA Tesla V100'. We use grid search to find the best hyperparameters for the models. We choose the embeddings dimensionality $d$ among \{50, 75, 100\}, the learning rate  among \{0.001, 0.01, 0.1\}, and the margin $\gamma$  among \{1, 5, 10\}. We train the models with a batch size of 10000 and a maximum of 500 epochs. The activation function for the CNN was $tanh(\cdot)$.  

\subsection{Empirical Results} 

We evaluate the performance of the models using Precision, PR-AUC and Accuracy. 
Table 2 and Table 3 show the performance of our proposed E-CLM and E-CLM++ models with the baselines. Results in bold font are the best obtained results. The goal of this work is to evaluate the models behind the Rakuten AIris TP. 

From Tables 2 and 3, we have the following findings: (1) E-CLM and E-CLM++ outperformed all other methods, using the experimental Ichiba, Travel and Tencent datasets on all metrics. E-CLM achieved 3.2\% and 10.7\% avg. improvement in PR-AUC metric on Ichiba and Travel datasets respectively over the Baseline TP. On the other hand, E-CLM++ achieved 3.8\% and 11.1\% avg. improvement in PR-AUC on Ichiba and Travel datasets, respectively. For precision and accuracy, E-CLM models also achieved the best performance. E-CLM++ worked well on Rakuten datasets; (2) Among TP+TransE, TP+ComplEX and TP+Graph Sage, though TP+Graph sage achieved better performance, but its performance is very close to TP+TransE which is surprising. We think TransE can work well when the data is very rich. However, complicated models might suffer from bias or overfitting problems; (3) The avg. performance of Rakuten Travel datasets are much higher than the Rakuten Ichiba datasets. MetaHeac \cite{zhu2021learning} developed by WeChat  achieved competitive results on Rakuten amd public datasets; (4) Though E-CLM is customized for Rakuten AIris TP, but it shows competitive performance for the public dataset as well. 


\subsection{Ablation Study}
We have introduced five different views for Rakuten Ichiba and Rakuten Travel related to target prospecting. Here, we present an ablation study to test the effectiveness of each view in E-CLM models. First, we show the results considering each view separately. Second, we show the combination of different views (please note total number of views are 5 so we included the most significant combination to give the readers a clear understanding). Please refer to Table 4 for the detail results. Major findings are: (1) Considering the performance of single views, Demography view and Ichiba view show more effectiveness than other views. (2) Family member view seems more effective for Rakuten travel than Rakuten Ichiba. (3) In the Travel dataset using Ichiba view can improve overall PR-AUC by 2.7\%. However, for Ichiba datasets, Travel view contributes 6.2\% improvement. 

\section{Conclusion}
This work introduces a novel lookalike algorithm that leverages a 360-degree view of customers along with the power of KG embedding. E-CLM models demonstrated the superior effectiveness in the experiments  compared to the other state-of-the-art approaches.  Here, we only showed the results of  two different services of  Rakuten. We believe that these results will inspire researchers to use cross service data for understanding users' different perspectives while building a lookalike model.
\section*{COMPANY PORTRAIT} 
\textbf{Rakuten Group, Inc.} is a leading global company that contributes to society by creating value through innovation and entrepreneurship. It operates in e-commerce, travel, digital content, communications, and Fintech sectors. Rakuten connects millions of customers to more than 90 different services with a wide range of options and felxibilities. Rakuten aims to build a fair society where individuals and companies are empowered to be successful in business and life.

\section*{Main Presenter} 
\textbf{Dr. Md Mostafizur Rahman} is a Lead Research Scientist in Rakuten Institute of Technology  (RIT) – the research division of Rakuten Group, Inc.  He develops machine learning algorithms to improve the customers’ experience on the various platforms offered by Rakuten and his recent focus is productionizing the graph based machine learning solutions developed by RIT. Previously, he was a Senior Engineer in Samsung Electronics Ltd.  He was a Visiting Researcher at the University of Queensland, Australia and worked as a Knowledge Graph Researcher in National Institute of Informatics, Japan.



\bibliographystyle{ACM-Reference-Format}
\bibliography{main}


\end{document}